\documentclass[preprint,12pt,authoryear]{elsarticle}
\usepackage{graphicx}
\usepackage{setspace}
\usepackage{amssymb}
\usepackage{amsmath}
\usepackage{float}
\usepackage[usestackEOL]{stackengine}
\usepackage{changepage}

\journal{Applied Soft Computing}

\begin{document}
\strutlongstacks{T}

\begin{frontmatter}



\title{Topic Modeling, Clade-assisted Sentiment Analysis, and Vaccine Brand Reputation Analysis of COVID-19 Vaccine-related Facebook Comments in the Philippines}


\author[inst1]{Jasper Kyle Catapang\corref{cor1}}
\cortext[cor1]{Corresponding author}
\affiliation[inst1]{organization={Department of English Language and Linguistics, University of Birmingham},
            state={Birmingham},
            country={United Kingdom}}

\author[inst2]{Jerome V. Cleofas}

\affiliation[inst2]{organization={Department of Sociology and Behavioral Sciences, De La Salle University},
            city={\\Manila City},
            country={Philippines}}

\begin{abstract}
Vaccine hesitancy and other COVID-19-related concerns and complaints in the Philippines are evident on social media. It is important to identify these different topics and sentiments in order to gauge public opinion, use the insights to develop policies, and make necessary adjustments or actions to improve public image and reputation of the administering agency and the COVID-19 vaccines themselves. This paper proposes a semi-supervised machine learning pipeline to perform topic modeling, sentiment analysis, and an analysis of vaccine brand reputation to obtain an in-depth understanding of national public opinion of Filipinos on Facebook. The methodology makes use of a multilingual version of Bidirectional Encoder Representations from Transformers or BERT for topic modeling, hierarchical clustering, five different classifiers for sentiment analysis, and cosine similarity of BERT topic embeddings for vaccine brand reputation analysis. Results suggest that any type of COVID-19 misinformation is an emergent property of COVID-19 public opinion, and that the detection of COVID-19 misinformation can be an unsupervised task. Sentiment analysis aided by hierarchical clustering reveal that 21 of the 25 topics extrapolated by topic modeling are negative topics. Such negative comments spike in count whenever the Department of Health in the Philippines posts about the COVID-19 situation in other countries. Additionally, the high numbers of laugh reactions on the Facebook posts by the same agency---without any humorous content---suggest that the reactors of these posts tend to react the way they do, not because of what the posts are about but because of who posted them.
\end{abstract}


\end{frontmatter}
\vspace{1em}

\section*{Highlights}
\begin{enumerate}
    \item COVID-19 vaccine hesitancy in the Philippines is primarily due to misinformation
    \item Posting about the COVID-19 situation in other countries increase negative comments
    \item Laugh reactions on humorless posts suggest they are targeted at the source of posts
    \item Sputnik V, AstraZeneca, and Sinovac suffer from a negative public reputation
    \item Ministries of health and stakeholders in vaccination campaigns are recommended to employ interventions that correct misinformation, engage people and use local narratives of success
\end{enumerate}

\section{Introduction}

The COVID-19 pandemic has drastically affected the overall wellness and health of the entire world. On January 30, 2020, the first case of COVID-19, in the Philippines, has been reported by the country's Department of Health (DOH). COVID-19 is a respiratory disease caused by the SARS-CoV-2 virus---first identified in the province of Wuhan, located in China \citep{paules_2020}.\\

Over a year later, on March 1, 2021, the Philippines started with its COVID-19 vaccination program. Vaccine hesitancy among Filipinos is an ongoing phenomenon that the national vaccine campaign efforts are struggling with \citep{alfonso_2021}. According to a recent study in the Philippines \citep{caple_2021}, only 62.5\% of their 7,193 respondents are willing to be vaccinated against COVID-19. A majority of the same respondents are only willing to be inoculated after many others have received the vaccine or after political figures have done so \citep{caple_2021}. Additionally, the participants' preferences of vaccine brand are also studied; 59.7\% of the participants are confident in a USA-made or European-made COVID-19 vaccine \citep{caple_2021}.\\

Sentiment analysis is a common natural language processing (NLP) task that has been done on a number of studies regarding COVID-19 public opinion \citep{melton_2021, garcia_2021}. It computationally classifies the polarity of text data---neutral, positive, or negative sentiment. This is primarily done since gauging the sentiment of the public, especially on critical topics such as a pandemic like COVID-19, help determine possible policies and interventions that could shape the actions that society takes. Furthermore, topic modeling is proposed as part of the NLP pipeline of the same studies mentioned earlier. Topic modeling is an unsupervised technique for obtaining the relevant ideas that public opinion holds, more of which is discussed in a later section of this article.\\

This paper proposes a pipeline for understanding the public opinion in the Philippines regarding COVID-19 and the country's vaccination efforts. The semi-supervised pipeline is named Vaccine-related Exploratory Research via Topic Extraction, Brand Reputation Analysis, and Topic Emotions (VERTEBRATE). It is comprised of three main modules: a topic extraction or topic modeling of opinions, an analysis of vaccine brand reputation, and a sentiment analysis of the same set of opinions. Topic modeling is performed through the use of contextual embeddings provided by a multilingual transformer architecture---discussed further in Section 2.3. The topics obtained are compared and associated with the vaccine brands available in the Philippines through cosine similarity. Next, a hierarchical clustering of the topics extrapolated through topic modeling is used to assign the labels used for sentiment analysis. Sentiment analysis on the data is performed by five different classification algorithms---assessing which architecture models the sentiments of the text data most effectively. The classification algorithms proposed are XGBoost, LightGBM, K-nearest neighbors, Naive Bayes' algorithm, and support vector machine. The public opinions considered for this study are comments made on 50 different Facebook posts by the official page of the Department of Health Philippines. The posts are part of the RESBAKUNA campaign of the same government department.


\section{Methodology}
A repository dedicated for this study---containing the link to the raw datasets, the source code of the different analyses performed in the study, and other miscellaneous files---can be found on Github.\footnote{https://github.com/jcatapang/COVID-19-VERTEBRATE}\\

\subsection{Data}
The data is comprised of around 100 top comments for each of the 50 Facebook posts by the official page of the Department of Health Philippines. These comments are primarily in English and Filipino. The 50 posts are the search results for querying the string: ``\#RESBAKUNA \#BIDASolusyon \#BIDAangMayDisiplina", and setting the year to 2021. The comments obtained range from April 20, 2021 up to September 9, 2021. These query parameters cover the entire call-for-vaccination campaign by the Department of Health Philippines until September 9, 2021. The ``top comments" sorting filter by Facebook is based on the number of reactions the comment has. A total of 4,877 comments is extracted from the 50 Facebook posts. In addition to the comments, the timestamps and the contents of the post they're commenting on are also extracted. The data is collected through Selenium and Python 3. Preprocessing of the Facebook comments are also done. The data scraped are preprocessed by removing several stop words, removing punctuations, removing emojis, and converting all letters to lowercase. This preprocessed dataset is the dataset utilized in the experiment.\\

\subsection{N-grams}
N-grams are subsequences, sized $n$, of virtually any sequence---like text and speech to name a few \citep{dai_2020}. In natural language processing, n-grams are more commonly used for word sequences. For example, ``vaccines work effectively" and ``the viral strains" are examples of trigrams (3-grams, n=3) of the text: ``COVID-19 vaccines work effectively on any of the viral strains." Word n-grams are used to model the co-occurring words within text data in order to find out the different, frequent combination of words that may help in modeling the data for any natural language processing task \citep{dai_2020}. In this study, word n-grams would be extracted from the Facebook comments to model the frequently occurring combinations of words in the data.\\

\subsection{Topic modeling}
Topic modeling provides a fast and effective unsupervised extraction of topics from text data that are subjective in nature \citep{melton_2021}. These texts include product and service reviews and social media posts. Topic modeling techniques such as Latent Dirichlet Allocation (LDA) and non-negative matrix factorization (NMF)---for long pieces of text---and biterm topic modeling---for short pieces of text, like tweets---have been the go-to algorithms of numerous studies utilizing topic models \citep{yan_2013}. However, recent advances in deep learning have enabled the integration of transformer architectures even for topic modeling \citep{abuzayed_2021}. The most popular transformer architecture, Bidirectional Encoder Representations from Transformers (BERT), provide context through its embeddings---due to its bidirectionality \citep{abuzayed_2021}. This paper leverages that extra layer of context to extrapolate high-quality topics through a multilingual BERT model that the Python library BERTopic provides \citep{grootendorst_2020}. A temporal variation of the topic model is also proposed to see the evolution of the topics with respect to time.\\

\subsection{Vaccine brand reputation analysis}
Public opinion regarding the COVID-19 pandemic has been the subject of multiple NLP studies \citep{melton_2021, lyu_2021, garcia_2021}. However, these studies---although mentioning vaccine-related topics and sentiments---have not discussed the public image of the different COVID-19 vaccine brands themselves, according to the extracted comments. In this article, cosine similarity of the BERT topic embeddings, obtained from the topic modeling experiment earlier, is proposed to associate the different COVID-19 vaccine brands to the different topics extracted from the Facebook comments. This technique is similar to the approach proposed by \citep{thongtan_2019}. The different COVID-19 vaccine brands found in the data are: Pfizer, Moderna, AstraZeneca, Johnson \& Johnson, Sputnik V, and Sinovac.\\

\subsection{Clade-assisted sentiment analysis}
One of the most used clustering techniques, hierarchical clustering is used to gain insights from the structure of a dataset. In this type of clustering, a pairwise measure of dissimilarity is used in order to assess the distance between two sets. This measure is called a linkage \citep{dogan_2021}. The linkage that BERTopic utilizes is Ward's linkage \citep{grootendorst_2020}. Ward's method is illustrated in Equation~\ref{wards_method}. The cluster distances initially used in Ward's method are thus defined to be the square of the Euclidean distance between the data points.

\begin{equation}
    d_{{ij}}=d(\{X_{i}\},\{X_{j}\})={\|X_{i}-X_{j}\|^{2}}.
    \label{wards_method}
\end{equation}

The hierarchy produced by Ward's method consists of sub-hierarchies named clades \citep{dogan_2021}. These clades of topics extracted by BERTopic is used to assign sentiments to the different topics. With these clades dictating the sentiment of different topics, the performance of the sentiment analysis done on the data relies on the quality of the output of the hierarchical clustering algorithm. In addition, this clade-assisted sentiment analysis has eliminated the need to manually label sentiments to each of the Facebook comments studied in the experiment, effectively making the proposed VERTEBRATE pipeline a semi-supervised learning approach.\\

For the purposes of this study, only the positive and negative sentiments are modeled. Sentiment classification is done by five different classification algorithms: XGBoost, LightGBM, K-nearest neighbors (KNN), Naive Bayes' algorithm, and support vector machine. These different models allow various approaches to model the data---some of which are expected to yield generally favorable results. XGBoost and LightGBM have their own respective Python libraries, namely \textit{xgboost} and \textit{lightgbm}, while the default models for KNN, Naive Bayes' algorithm, and support vector machine provided by \textit{scikit-learn} are used for the series of experiments. In terms of the data to be classified, further steps are proposed to optimize the results of the classification task. First, the neutral topics of the dataset are removed. The positive clade or clades assign a positive sentiment to the comments belonging to the same clade---the same process is done for the negative sentiments. Next, the dataset is balanced before splitting it into training and test sets through oversampling. Training data used for sentiment analysis is 80\%. 560 positive comments and 560 negative comments are used for training, while a separate set of 140 positive comments and 140 negative comments are used for testing. The texts found in the training and testing datasets are vectorized via term-frequency inverse document frequency or TF-IDF.\\

\subsection{Evaluation metrics}
The standard evaluation metrics for classification tasks are accuracy, precision, recall, and the F1 score. These metrics are basic measures obtainable from the confusion matrix of a machine learning model. These evaluation metrics are not enough to provide a reliable measure of correctness of the representations learned by the variables measured. This can be properly calculated by Cohen's kappa statistic \citep{mchugh_2012}. More precisely, Cohen's kappa statistic is used to test interrater reliability. This statistic is proposed by Cohen since percent agreement cannot account for chance agreement. This statistic, shown in Equation~\ref{kappa}, ranges from -1 to 1. In Equation~\ref{kappa}, $P_o$ refers to the relative observed agreement among raters, while $P_e$ is the chance agreement's hypothetical probability. In this study, the ground truth is provided by the assignments made by the clades found in the hierarchical clustering model, and the other rater is the classification output.

\begin{equation}
    \kappa = \frac{P_o - P_e}{1 - P_e}.\label{kappa}
\end{equation}

According to \citet{mchugh_2012}, the interpretation of Cohen's kappa statistic is enumerated in Table~\ref{cohen}. For this study, a strong or almost perfect level of agreement is favorable for each classifier.\\

\begin{table}[H]
\centering
\begin{tabular}{|l|l|l|}
\hline
{$\boldsymbol\kappa$} & \textbf{level of agreement} & \textbf{\% of reliable data} \\ \hline
0–.20                                       & none                                            & 0–4\%                                                     \\ \hline
.21–.39                                     & minimal                                         & 4–15\%                                                    \\ \hline
.40–.59                                     & weak                                            & 15–35\%                                                   \\ \hline
.60–.79                                     & moderate                                        & 35–63\%                                                   \\ \hline
.80–.90                                     & strong                                          & 64–81\%                                                   \\ \hline
$>$ .90                                    & almost perfect                                  & 82–100\% \\ \hline                                                
\end{tabular}
\caption{Interpretation of Cohen's kappa statistic}
\label{cohen}
\end{table}

\section{Results}
\subsection{N-gram frequencies}
The n-gram frequencies, discussed in Section 2.2, are extracted to obtain the most commonly occurring sets of words in the data. Table~\ref{unigram} contains the top 10 frequent unigrams that are also relevant to the study of COVID-19 and its vaccine.\\

\begin{table}[H]
    \centering
    \begin{tabular}{ | c | c | }
        \hline
        \textbf{unigram} & \textbf{frequency} \\ \hline
        vaccine & 1541 \\ \hline
        \textit{bakuna} & 1249 \\ \hline
        \textit{wala} & 623 \\ \hline
        \textit{tao} & 555 \\ \hline
        DOH & 545 \\ \hline
        dose & 499 \\ \hline
        vaccinated & 482 \\ \hline
        \textit{sana} & 384 \\ \hline
        \textit{ayaw} & 335 \\ \hline
        \textit{namatay} & 318 \\ \hline
    \end{tabular}
    \caption{Top 10 most common relevant unigrams}
    \label{unigram}
\end{table}

The unigrams in Table~\ref{unigram} are the following: ``vaccine", ``bakuna", ``wala", ``tao", ``DOH", ``dose", ``vaccinated", ``sana", ``ayaw" and ``namatay". ``Bakuna" is ``vaccine" in Filipino. ``Wala" is a Filipino word used to describe the absence of something. In this context it can pertain to the vaccine or other COVID-19 essentials. ``Tao" is ``person" in Filipino. ``Sana" is a Filipino expression of hope or longing. ``Ayaw" is a Filipino word that signals disapproval or rejection. ``Namatay" is a Filipino word for the dead (as a noun) or died (as a past tense verb).\\

\begin{table}[H]
    \centering
    \begin{tabular}{ | c | c | }
        \hline
        \textbf{bigram} & \textbf{frequency} \\ \hline
        of, health & 198 \\ \hline
        department, of & 195 \\ \hline
        health, philippines & 185 \\ \hline
        2nd, dose & 166 \\ \hline
        fully, vaccinated & 123 \\ \hline
        the, vaccine & 118 \\ \hline
        1st, dose & 103 \\ \hline
        second, dose & 96 \\ \hline
        side, effects & 83 \\ \hline
        first, dose & 78 \\ \hline
    \end{tabular}
    \caption{Top 10 most common relevant bigrams}
    \label{bigram}
\end{table}

Table~\ref{bigram} contains the ten most common bigrams that are relevant to the study. The bigrams in the table are self-explanatory except for the first three. The bigrams ``of health", ``department of", and ``health philippines" are parts of the ``Department of Health Philippines".\\

\begin{table}[H]
    \centering
    \begin{tabular}{ | c | c | }
        \hline
        \textbf{trigram} & \textbf{frequency} \\ \hline
        department, of, health & 193 \\ \hline
        of, health, philippines & 185 \\ \hline
        to, be, vaccinated & 68 \\ \hline
        to, get, vaccinated & 38 \\ \hline
        in, the, philippines & 22 \\ \hline
        want, to, be & 20 \\ \hline
        vaccine, is, not & 18 \\ \hline
        be, vaccinated, but & 17 \\ \hline
        don't, want, to & 16 \\ \hline
        \textit{wag, pilitin, ayaw} & 15 \\ \hline
    \end{tabular}
    \caption{Top 10 most common relevant trigrams}
    \label{trigram}
\end{table}

Table~\ref{trigram} lists the ten most frequent and relevant trigrams. Several trigrams extracted from the data require further discussion. The first two trigrams are a similar case to the explanation made earlier about being parts of the ``Department of Health Philippines". The trigram ``want to be" expresses desire but it cannot stand on its own without being part of a 4-gram. Specifically, ``want to be" needs to be proceeded by another word. The trigram ``be vaccinated but", expresses concern or reservation to being vaccinated. Lastly, ``wag pilitin ayaw" roughly translates to ``don't force someone who's unwilling".\\

For the last n-gram frequencies, Table~\ref{fourgram} illustrates the 4-grams present in the data. ``Department of health philippines", ``of health philippines department", ``health philippines department of", and ``philippines department of health" are all variants of Department of Health Philippines. As discussed earlier, the trigram ``want to be" requires another word after it. The 4-gram ``want to be vaccinated" completes the thought. The rest of the 4-grams require no explanation.\\

\begin{table}[H]
    \centering
    \begin{tabular}{ | c | c | }
        \hline
        \textbf{4-gram} & \textbf{frequency} \\ \hline
        department, of, health, philippines & 185 \\ \hline
        to, be, vaccinated, but & 17 \\ \hline
        want, to, be, vaccinated & 15 \\ \hline
        of, health, philippines, department & 11 \\ \hline
        health, philippines, department, of & 11 \\ \hline
        philippines, department, of, health & 11 \\ \hline
        want, to, get, vaccinated & 9 \\ \hline
        don't, want, to, be & 9 \\ \hline
        god, bless, us, all & 9 \\ \hline
        face, mask, face, shield & 9 \\ \hline
    \end{tabular}
    \caption{Top 10 most common relevant 4-grams}
    \label{fourgram}
\end{table}

\subsection{BERT topic model}
\subsubsection{Topics}
\begin{table}[H]
    \centering
    \begin{adjustwidth}{-0.5cm}{}
    \begin{tabular}{ | c | c | }
        \hline
        \textbf{topic name} & \textbf{terms} \\ \hline
        vaccination & \Centerstack{vaccine, vaccinated, is, are, we,\\vaccines, vaccination, have, dose, if}\\ \hline
        vaccine dose scheduling & \Centerstack{dose, 2nd, second, 1st, ulo,\\lng, araw, nabakunahan, doses, nung}\\ \hline
        vaccine registration & \Centerstack{online, bakuna, nga, tao, registration,\\yan, magpabakuna, register, kaso, nagpabakuna}\\ \hline
        vaccine deaths & \Centerstack{maraming, talaga, namatay, pwede, namamatay,\\tao, natin, ilang, sasabihin, saudi}\\ \hline
        \Centerstack{department of health\\negative comments} & \Centerstack{philippines, health, pandemic, ineptitude, firing,\\naideliver, muncipyo, philippinnes, konsulta, nanamatay}\\ \hline
        Belgium & \Centerstack{belgium, population, people, respect, their,\\negative, million, propaganda, social, news}\\ \hline
    \end{tabular}
    \end{adjustwidth}
    \caption{Topics 1 to 6}
    \label{topics_1-6}
\end{table}

After using a multilingual BERT model for topic modeling on the 4,877 comments, the 25 topics that the algorithm produces are assigned topic names by a health professional by assessing the terms associated to the topic cluster and its representative comment. Table~\ref{topics_1-6} describes topics 1 to 6 out of 25. Topic 1 talks about vaccination in general. A representative comment reads: \textit{``everyday on fb i see people passing away so sad that you're doing everything to push it to the people"}. Topic 2 talks about the scheduling of the vaccines' 1st and 2nd doses. This is a representative comment for Topic 2: \textit{``got my first dose last friday and i'm okay so please do your part on protecting yourself and the community"}. Topic 3 is about vaccine registration. A representative comment from Topic 3 reads: \textit{``bakt cozn q nagpabakuna bakt laging sumasama  pkiramdam absent trabaho dati wala nmn syang naramdamang sakt oh bakt"}. Roughly translated, the representative comment means: \textit{``Why did my cousin feel unwell after getting vaccinated? He/she didn't go to work because of it. He didn't feel any symptoms or illnesses beforehand. Why is that?"} On the other hand, Topic 4 purports alleged deaths caused by or linked to the COVID-19 vaccines. A representative comment tells a story about Topic 4: \textit{``Buti pa kayo may kwento samantala ung kababayan ko na pupunta sana ng Saudi namatay pagkatapos bakunahan."} When translated it reads: \textit{``A person in my town, that was supposedly going to Saudi, died after getting vaccinated."} Next, Topic 5 clusters comments that talk about the Department of Health Philippines and the negative comments about the ministry. Here is a representative comment for the same cluster: \textit{``Affected by DOH's ineptitude during the Covod pandemic: over a 100m Filipinos"}. Meanwhile, Topic 6 pertains to the comments made on the Facebook post by the Department of Health Philippines about Belgium. An excerpt from the post reads: "Belgium chose to get vaccinated, and they are reaping the rewards! The results? No more lockdowns. Reopened cinemas. Reopened restaurants and bars." A representative comment reads: \textit{``Belgium.. 12 million people... Philippines.. 110 million... Its not really the same"}.\\

\begin{table}[H]
    \centering
    \begin{adjustwidth}{-0.25cm}{}
    \begin{tabular}{ | c | c | }
        \hline
        \textbf{topic name} & \textbf{terms} \\ \hline
        healthy lifestyle & \Centerstack{philippines, health, vitamins, exercise, filipino,\\filipinos, foods, why, healthy, we}\\ \hline
        vaccine complaints & \Centerstack{bakuna, ninyo, tao, lalo, mabakunahan,\\niyo, magpabakuna, yan, namatay, bayan}\\ \hline
        vaccine deaths & \Centerstack{bakuna, tao, kyo, namatay, nyong,\\lalo, ninyo, nio, naba, kng}\\ \hline
        \Centerstack{vaccine requests and\\vaccine complaints} & \Centerstack{bakuna, pinas, bakunado, mgpabakuna, magpabakuna,\\madami, maherap, magpa, bulacan, namamatay}\\ \hline
        dismissal/acceptance & \Centerstack{kau, basta, sakin, niyo, bumaba,\\sainu, manlang, lakad, mangyari, bakunahan}\\ \hline
        vaccine concerns & \Centerstack{tao, buhay, talaga, kyo, yan,\\di, puh, gano, yun, nalang}\\ \hline
    \end{tabular}
    \end{adjustwidth}
    \caption{Topics 7 to 12}
    \label{topics_7-12}
\end{table}

Table~\ref{topics_7-12} enumerates Topics 7 to 12 of the BERT topic model. Topic 7 might sound like having a healthy lifestyle is a topic promoting complete wellness; this is not the case. Upon manual inspection, comments that belong to Topic 7 talk about people's preference of vitamins, exercise, and healthy food over getting inoculated and ultimately dismissing the COVID-19 vaccine as important or necessary. A representative comment of this cluster reads: \textit{``Dami nga complaint sainyo ipaliwanag nyo yan..pnpilit kc n mgpabakuna..kawawa ung nauto..instead bakuna dapat mga vitamins at masustansyang foods ang ibinibigay"}. This is translated as: \textit{``There's a lot of complaints about you. Explain this. You're forcing people to get vaccinated. Those gullible people are pitiful. Instead of vaccines, you should give out vitamins and nutritious foods."} Topic 8 talks about complaints about the COVID-19 vaccines. Here is a representative comment for the cluster: \textit{``Kaya nman talaga ang problem kulang sa gamot dami gusto pa bakuna"}. When translated, it reads: \textit{``It is certainly doable. The problem is the lack of supplies. There are a lot of people who want to receive the vaccine."} Topic 9 is similar to Topic 4 of Table~\ref{topics_1-6}. Topic 10 is similar to Topic 8. Meanwhile, the next topic contains comments about dismissing the virus and accepting their fate or futility of government efforts. A representative comment from Topic 11 reads: \textit{``Yung para sa akin po bigay ko na sa mga mas nangangailangan..bahala na mag ka covid ako basta kau hindi...mahal ko kau.."}. It roughly translates to \textit{``I'll just give away my slot for the vaccine for those who need it more. I leave it up to chance if I contract COVID-19, so long as you guys don't. I love you."} The last topic for Table~\ref{topics_7-12}, this cluster talks about concerns regarding the COVID-19 vaccines. Here is a representative comment about Topic 12: \textit{``Utang na loob, DOH, magbasa po kayo ng comments. Hayaan niyong tumagos sa mga puso niyo ang hinanaing at hinagpis ng tao!"} Translated, the representative comment means: \textit{``For crying out loud, DOH. Read the comments. Let the cries and resentment of the people pierce through your hearts!"}\\

\begin{table}[H]
    \centering
    \begin{adjustwidth}{-1.75cm}{}
    \begin{tabular}{ | c | c | }
        \hline
        \textbf{topic name} & \textbf{terms} \\ \hline
        vaccine efficacy on covid-19 variants & \Centerstack{virus, variant, nabakunahan, bakuna, pandemya,\\yan, pandemic, pangangalakal, sabay, niyo}\\ \hline
        religious comments & \Centerstack{god, he, trust, will, him,\\life, faith, his, jesus, believe}\\ \hline
        health workers & \Centerstack{hospital, doctor, doktor, medical, health,\\doctors, hyperthyroidism, sinofarm, nadiguk, namatay}\\ \hline
        department of health and workforce & \Centerstack{department, health, philippines, workers, agency,\\employer, essential, companies, priority, manila}\\ \hline
        vaccine requirements and registration & \Centerstack{schedule, house, priority, mandatory, bakuna,\\list, requirements, required, registered, register}\\ \hline
        vaccine trials, side effects, and testing & \Centerstack{clinical, trial, trials, effects, test,\\swab, bayad, effect, tested, testing}\\ \hline
    \end{tabular}
    \end{adjustwidth}
    \caption{Topics 13 to 18}
    \label{topics_13-18}
\end{table}

Table~\ref{topics_13-18} shows Topics 13 to 18 of the BERT topic model. Topic 13 consists of comments about the efficacy of the vaccines on the COVID-19 variants. A representative comment on the subject reads: \textit{``Sinasabi nyo nag upgrade na ang virus.Delta na at baka sa susunod na mag upgrade ang virus charlie na tapos ang susunod bravo na pero bakuna ninyo hindi nag upgrade."} The translated comment is as follows: \textit{``You're saying the virus upgraded. It's Delta already, and next time it's going to be Charlie, Bravo, etc., but your vaccines don't upgrade along with the virus."} Topic 14 contains religious comments and require no further elaboration. Health workers are the topic of Topic 15. A comment from the cluster reads: \textit{``andaming mas matatalino dito kesa sa mga doctor"}. When translated: \textit{``There are a lot more people that are smarter than doctors"}. Topic 16 talks about DOH and the workforce. Here is a comment from the same topic: \textit{``Puro senior nmn inuuna nyo panu ung mg labas pasok ng bahay lalo n kmi ngtatrabaho ..kht san n lng senior ang priority ..gumagala ata mga senior ngaun kaya cla lagi priority.."}. In English, it reads: \textit{``You're only prioritizing seniors. How about those who keep on going outside, especially us with jobs? Everywhere you go seniors are the only priority. Maybe seniors just keep on dilly-dallying, that's why."} Vaccine requirements and registration are the focus of the cluster of comments found in Topic 17. A comment from the cluster reads: \textit{``Ay naku hindi n lng ako aalis kung yan requirements"}. In English, it means: \textit{``Oh boy! I'm not leaving anymore if those are the requirements."} The last topic in Table~\ref{topics_13-18} is about vaccine trials, side effects, and testing. A comment from Topic 18 reads: \textit{``Dami na namatay sa bakuna .. still on clinical trial"}. When translated, the comment means: \textit{``Many died because of the vaccine. It's still in its clinical trials."}\\

\begin{table}[H]
    \centering
    \begin{adjustwidth}{-0.75cm}{}
    \begin{tabular}{ | c | c | }
        \hline
        \textbf{topic name} & \textbf{terms} \\ \hline
        religious comments & \Centerstack{lord, jesus, god, pray, amen,\\mary, godbless, prayer, blessed, faith}\\ \hline
        vaccine and illnesses & \Centerstack{sakit, sugat, bakuna, gamot, nigusyo,\\mabibilhin, healasone, pagkatiwalaan, ferandoz, shara}\\ \hline
        vaccine side effects & \Centerstack{effect, effects, magpabakuna, katawhan, kayong,\\safe, adverse, epekto, bad, lalamig}\\ \hline
        vaccine blaming & \Centerstack{visa, bakuna, tao, ngaun, magpabakuna,\\malay, lalo, international, yan, governo}\\ \hline
        face masks and face shields & \Centerstack{face, mask, shield, facemask, faceshield,\\magface, swab, alcohol, pina, video}\\ \hline
        vaccine hesitancy & \Centerstack{safe, lockdown, hinde, yan, bakuna,\\talga, talaga, waiver, quarantine, kasali}\\ \hline
        \Centerstack{prioritization of vaccine\\administration} & \Centerstack{senior, seniors, citizen, citizens, barangay,\\magpabakuna, campaign, priority, month, apply}\\ \hline
    \end{tabular}
    \end{adjustwidth}
    \caption{Topics 19 to 25}
    \label{topics_19-25}
\end{table}

The last table of topics, Table~\ref{topics_19-25}, lists down familiar topics from previous tables. All topics except face masks and face shields are similar topics to the ones discussed in earlier tables. Topic 23, the comments on face masks and face shields, contains different preventive measures. Face masks and shields comprise the majority of the cluster. A representative comment reads: \textit{``Kung ligtas na sa vaccine huwag na mag suot ng face mask at face shield"}. In English, it reads: \textit{``If the vaccine protects you already, there's no need to wear a face mask and a face shield."}\\

\subsubsection{Percentage distribution}
\begin{table}[H]
\begin{adjustwidth}{0.5cm}{}
\begin{tabular}{|l|l|l|}
\hline
\textbf{topic name}                       & \textbf{comments} & \textbf{percentage} \\ \hline
vaccination                               & 1895                        & 38.85585            \\ \hline
outlier                                   & 1357                        & 27.82448            \\ \hline
vaccine dose scheduling                   & 168                         & 3.444741            \\ \hline
vaccine registration                      & 100                         & 2.050441            \\ \hline
vaccine deaths                          & 96                          & 1.968423            \\ \hline
department of health negative comments    & 91                          & 1.865901            \\ \hline
Belgium                                   & 89                          & 1.824892            \\ \hline
healthy lifestyle                         & 88                          & 1.804388            \\ \hline
vaccine complaints                        & 86                          & 1.763379            \\ \hline
vaccine deaths                          & 84                          & 1.72237             \\ \hline
vaccine requests and vaccine complaints   & 77                          & 1.578839            \\ \hline
dismissal/acceptance                      & 75                          & 1.537831            \\ \hline
vaccine concerns                          & 71                          & 1.455813            \\ \hline
\end{tabular}
\end{adjustwidth}
\caption{Top 13 topic percentage distribution}
\label{top13}
\end{table}

The percentage distribution of the topics in the topic model is illustrated in Tables~\ref{top13} and~\ref{bottom12}. The general comments on vaccination and outlier comments dominate the data, making up 38.86\% and 27.82\% of the data, respectively. Together, these two topics are considered to be of indeterminable sentiments, and is not included in the clade-assisted sentiment analysis in Section 3.5.\\

\begin{table}[H]
\begin{adjustwidth}{0.5cm}{}
\begin{tabular}{|l|l|l|}
\hline
\textbf{topic name}                       & \textbf{comments} & \textbf{percentage} \\ \hline
vaccine efficacy on covid-19 variants     & 63                          & 1.291778            \\ \hline
religious comments                      & 59                          & 1.20976             \\ \hline
department of health and workforce        & 56                          & 1.148247            \\ \hline
health workers                            & 56                          & 1.148247            \\ \hline
vaccine requirements and registration     & 50                          & 1.02522             \\ \hline
vaccine trials, side effects, and testing & 47                          & 0.963707            \\ \hline
religious comments                      & 46                          & 0.943203            \\ \hline
vaccine and illnesses                     & 45                          & 0.922698            \\ \hline
vaccine side effects                      & 41                          & 0.840681            \\ \hline
vaccine blaming                           & 39                          & 0.799672            \\ \hline
face masks and face shields               & 35                          & 0.717654            \\ \hline
vaccine hesitancy                         & 33                          & 0.676645            \\ \hline
prioritization of vaccine administration  & 30                          & 0.615132            \\ \hline
\end{tabular}
\end{adjustwidth}
\caption{Bottom 12 topic percentage distribution}
\label{bottom12}
\end{table}

\subsubsection{Hierarchical clustering}
\begin{figure}[H]
    \hspace*{-2cm}                                                           
    \includegraphics[scale=0.5]{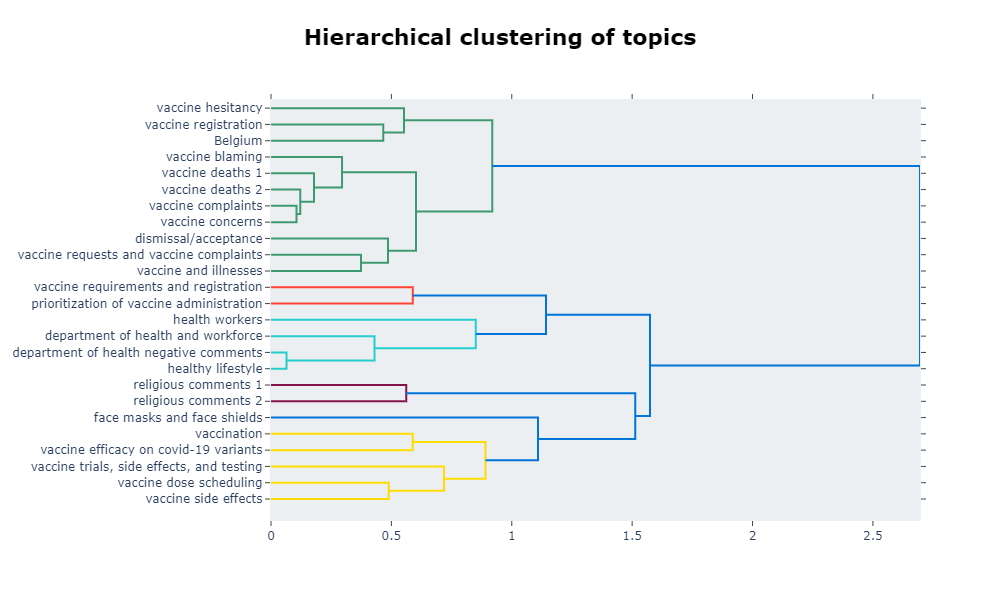}%
    \hspace{2mm}%
    \caption{Dendrogram of the BERT topic embeddings}
    \label{hierarchy}
\end{figure}

After constructing a dendrogram of the BERT topic embeddings, the hierarchy of the topics found within the data is demonstrated in Figure~\ref{hierarchy}. The topics of indeterminable sentiments are removed as they skew the data. As shown in Figure~\ref{hierarchy}, the data is divided into two superclades. The green superclade talks primarily about COVID-19 vaccine misinformation. The blue superclade is a combination of COVID-19 misinformation, frequently asked questions, and comments indicating desperation.

The red clade talks about the vaccination process itself. It consists of comments about the requirements for getting vaccinated, the registration process, and the priority classes that the government has implemented. The cyan clade talks about people---health workers, the workforce---along with complaints about the Department of Health Philippines, and people asking the ministry to give out vitamins and healthy food over the vaccines. The purple clade talks about religious comments. It is the only clade that have positive sentiments. In the clade-assisted sentiment analysis that is proposed in this study, the purple clade represents all training and testing data for the positive sentiment. The yellow clade can be considered the clade for frequently asked questions (FAQs). Lastly, the face masks and face shields topic stands on its own and makes a lone member clade.\\

\subsection{Temporal BERT topic model}
\begin{figure}[H]
    \hspace*{-2cm}                                                           
    \includegraphics[scale=0.45]{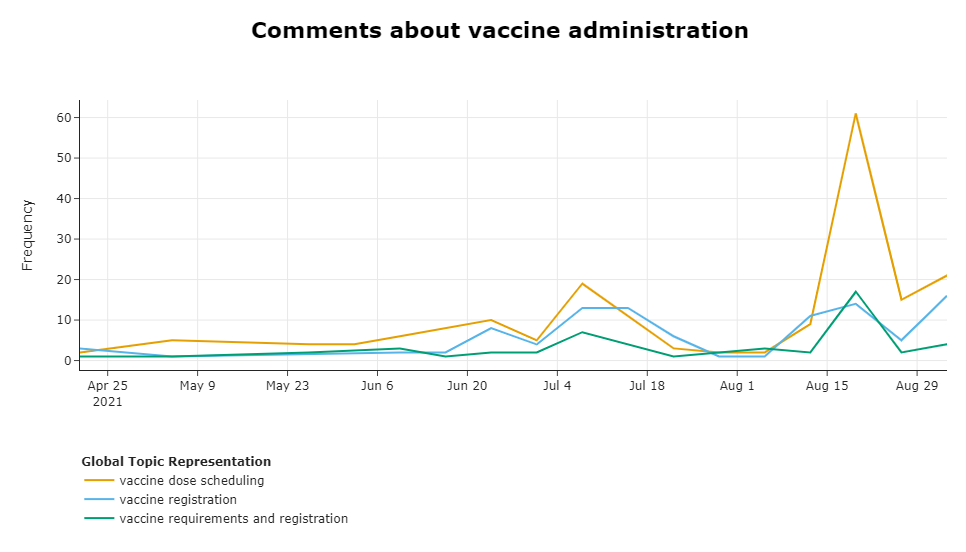}%
    \hspace{2mm}%
    \caption{Temporal distribution of vaccine administration comments}
    \label{vaccine-administration}
\end{figure}

The temporal distribution of the comments on vaccine dose scheduling, vaccine registration, vaccine requirements and registration can be seen in Figure~\ref{vaccine-administration}. A spike in the number of comments days after July 4, 2021 is observable. These are comments that were made when the Department of Health Philippines released posts encouraging people to get registered and get vaccinated. One post is about Minnesota's vaccination efforts dated July 6, 2021. Another content released by the Department of Health Philippines about Scotland's vaccination efforts was posted on July 7, 2021. Lastly, Belgium's vaccination efforts were also posted on July 8, 2021. Another spike in the number of comments occurred days after August 15, 2021. This is after two posts of the Department of Health Philippines went viral. One post reads: \textit{``What is your perception on the COVID-19 vaccines? Whether or not you have already been vaccinated, we’d like to hear from you! To access the survey, you may scan the QR code or click on this link: bit.ly/COVIDvaccinesurvey. The survey will be open from 20 August 2021 until 27 August 2021 or until the target number of respondents per region is reached. This survey ensures that any personal identifying information will be kept confidential. Thank you!"} This was posted on August 20, 2021. Another post invites people to get vaccinated since only 0.23\% of the vaccinated developed adverse reactions. The post was released on August 20, 2021 also.\\

\begin{figure}[H]
    \hspace*{-2cm}                                                           
    \includegraphics[scale=0.45]{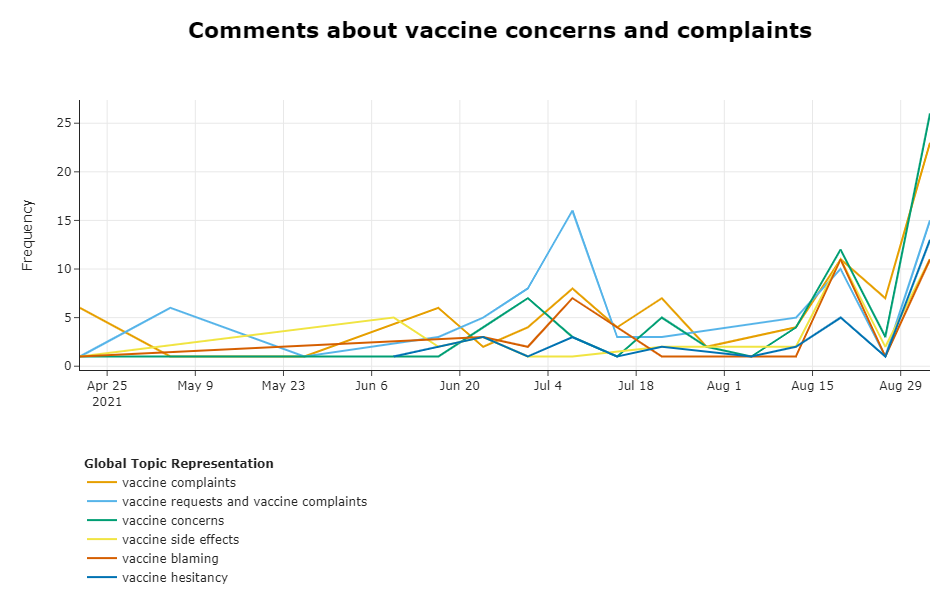}
    \hspace{2mm}%
    \caption{Temporal distribution of comments about vaccine concerns and complaints}
    \label{vaccine-concerns}
\end{figure}

The temporal distribution of the comments on vaccine complaints, vaccine requests and vaccine complaints, vaccine concerns, vaccine side effects, vaccine blaming, and vaccine hesitancy follow the same spikes as the previous temporal distribution of topics. The first spike in the distribution found in Figure~\ref{vaccine-concerns} is much abrupt when it comes to vaccine requests and vaccine complaints. They react to the same posts by DOH Philippines as before. The same spike can also be observed days after August 15, 2021. This spike is still caused by the two viral Facebook posts enumerated earlier. \\

After experience a drop in comments about vaccine concerns and complains before August 29, 2021, the comments suddenly rose drastically towards the month of September 2021. This is due to the continued appearance of Facebook posts by DOH Philippines that are highly reacted to and highly shared. One post about vaccine hesitant individuals garnered 26,000 total reactions as of September 9, 2021. 2,500 reactions of which, are laugh reactions. The post was released on September 1, 2021. Another post talking about victims of COVID-19 misinformation was posted on September 5, 2021. As of September 9, 2021, the post has a total of 14,000 reactions, 1,400 of which are laugh reactions. As the last example, a post released on September 7, 2021 garnered 13,000 reactions as of September 9, 2021. 1,500 of the reactions are laugh reactions. The post is about encouraging loved ones to get vaccinated especially when they divert the topic to something else.\\

\begin{figure}[H]
    \hspace*{-2cm}                                                           
    \includegraphics[scale=0.45]{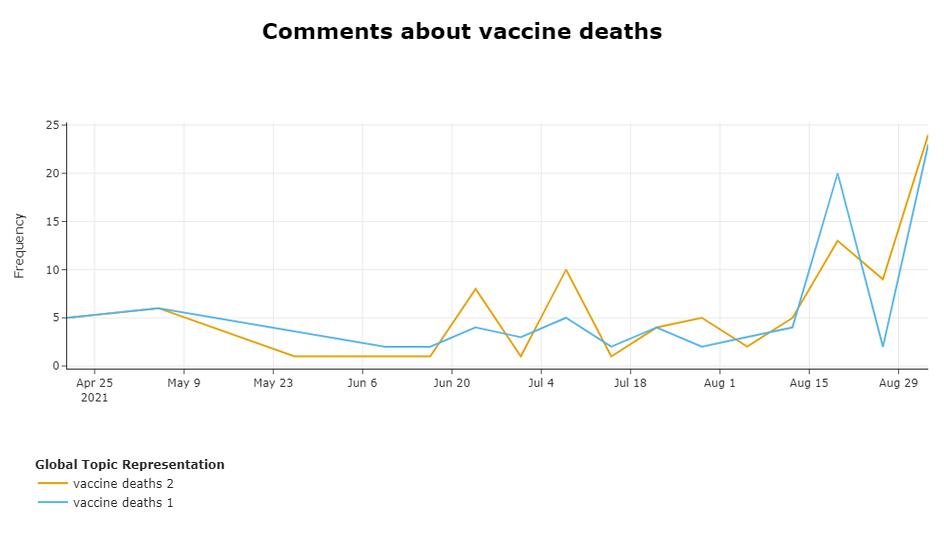}
    \hspace{2mm}%
    \caption{Temporal distribution of comments about alleged deaths caused by the vaccine}
    \label{vaccine-deaths}
\end{figure}

Figure~\ref{vaccine-deaths} focuses on the trend of comments about vaccine-related deaths through time. The spikes follow the trends explained in the earlier figures (see Figures~\ref{vaccine-administration} and~\ref{vaccine-concerns}). These comments occur around 25 times at most throughout the period, compared to the occurrences of comments regarding vaccine dose scheduling found in Figure~\ref{vaccine-administration}---peaking at 60 occurrences.

\subsection{Vaccine brand topic similarity}
\begin{figure}[H]
    \hspace*{-2cm}                                                           
    \includegraphics[scale=0.45]{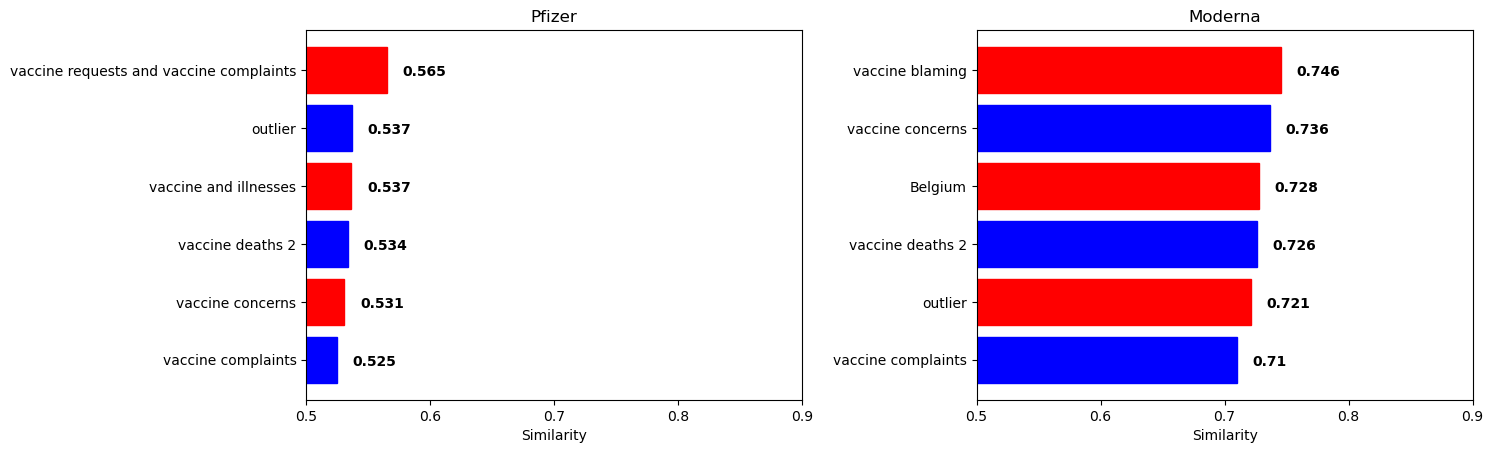}%
    \hspace{2mm}%
    
    \hspace*{-2cm}                                                           
    \includegraphics[scale=0.45]{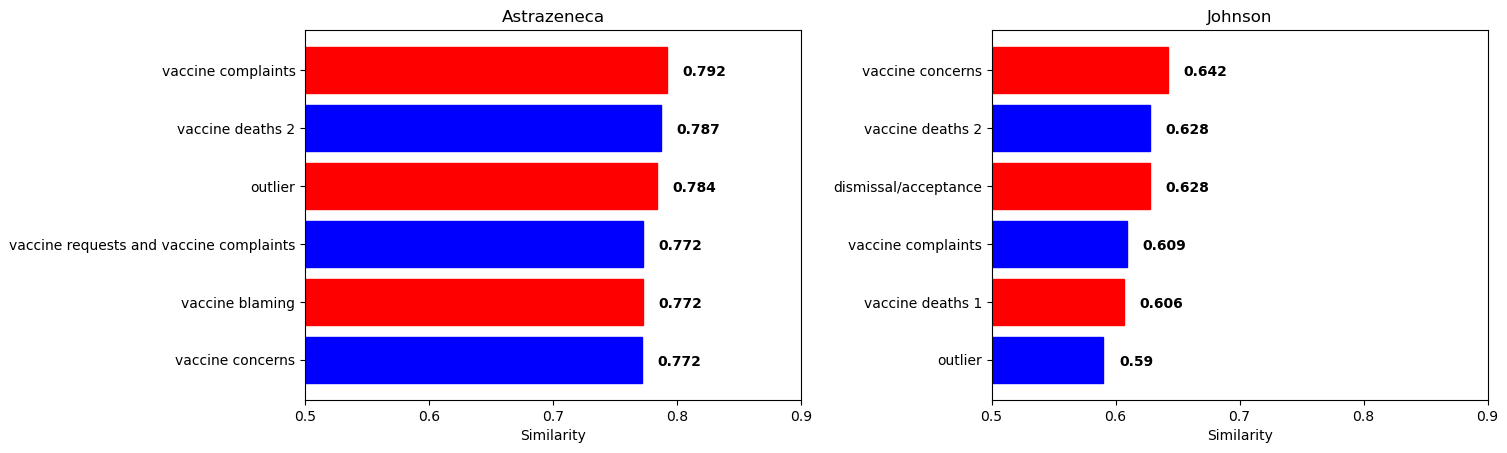}%
    \hspace{2mm}%

    \hspace*{-2cm}                                                           
    \includegraphics[scale=0.45]{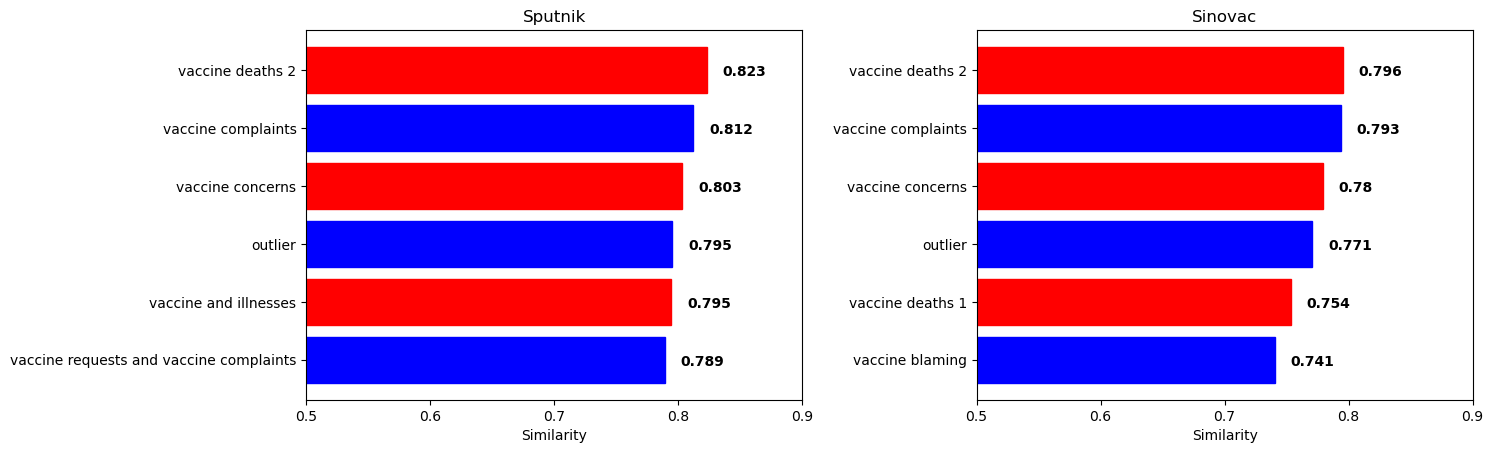}%
    \hspace{2mm}%
    \caption{Topic similarity for the different vaccine brands}
    \label{topic_similarity}
\end{figure}

Figure~\ref{topic_similarity} illustrates the topics similar to the vaccine brands according to the cosine similarity of their BERT topic embeddings. Pfizer has the lowest percentages of similarities to the BERT topics. Pfizer is associated with vaccine requests and vaccine complaints, vaccine and illnesses, vaccine deaths, vaccine concerns, and vaccine complaints. The similarity percentages can be found in the first subplot of Figure~\ref{topic_similarity}. Next, Moderna is associated with vaccine blaming, with a similarity of 74.6\%, vaccine concerns at 73.6\%, 72.8\% with the Belgium comments, 72.6\% with comments about vaccine-related deaths, and 71\% with vaccine complaints. AstraZeneca is associated with vaccine complaints at 79.2\%, 78.7\% with vaccine deaths, and 77.2\% each for the three: vaccine requests and complaints, vaccine blaming, and vaccine concerns. The Johnson \& Johnson vaccine is associated with vaccine concerns at 64.2\%, vaccine deaths and dismissal/acceptance, both at 62.8\%, 60.9\% with vaccine complaints, and 60.6\% with vaccine deaths. The Sputnik V vaccine is the most complained about---largely due to supply shortages. It is associated with vaccine deaths at 82.3\% similarity, vaccine complaints at 81.2\%, 80.3\% with vaccine concerns, 79.5\% with vaccine and illnesess, and 78.9\% with vaccine requests and vaccine complaints. Sinovac is associated with a topic on vaccine deaths at 79.6\% similarity, 79.3\% with vaccine complaints, 78\% with vaccine concerns, another topic on vaccine deaths at 75.4\% similarity, and 74.1\% similarity with vaccine blaming. As previously mentioned, the Sinopharm vaccine is never mentioned in the data.\\

\subsection{Sentiment classification based on clade membership}
\begin{table}[H]
\centering
\begin{tabular}{|l|l|l|l|l|l|}
\hline
\textbf{seed}    & \textbf{accuracy} & \textbf{precision} & \textbf{recall} & \textbf{f1 score} & \textbf{$\boldsymbol\kappa$} \\ \hline
15 & 0.861 & 0.879 & 0.861 & 0.859 & 0.721 \\ \hline
27 & 0.879 & 0.888 & 0.879 & 0.878 & 0.757 \\ \hline
32 & 0.946 & 0.948 & 0.946 & 0.946 & 0.893 \\ \hline
45 & 0.932 & 0.933 & 0.932 & 0.932 & 0.864 \\ \hline
51 & 0.939 & 0.942 & 0.939 & 0.939 & 0.879 \\ \hline
\textbf{AVERAGE} & 0.911 & 0.918 & 0.911 & 0.911 & 0.823 \\ \hline
\end{tabular}
\caption{Performance metrics of XGBoost on clade-assisted sentiment analysis}
\label{xgboost}
\end{table}

The results of the evaluation metrics on the XGBoost model are shown in Table~\ref{xgboost}. The XGBoost model manages to attain a minimum of 86.1\% and a maximum of 94.6\% accuracy across five seeds. The average accuracy of the model is 91.1\%. The precision's range is 0.879 up to 0.948, with an average of 0.918. The recall's range is 0.861 - 94.6 and average recall of 0.911. The F1 scores across five seeds range from 0.859 up to 0.946, with an average of 0.911. The model also has a relatively moderate minimum value of Cohen's kappa statistic, with a minimum of 0.721. Its maximum reliability---through Cohen's kappa statistic---reached 0.893, which is considered to approach a level of agreement that is almost perfect according to Table~\ref{cohen}. The average value of $\kappa$ of the XGBoost model is 0.823, which indicates a strong level of agreement.\\

\begin{table}[H]
\centering
\begin{tabular}{|l|l|l|l|l|l|}
\hline
\textbf{seed}    & \textbf{accuracy} & \textbf{precision} & \textbf{recall} & \textbf{f1 score} & \textbf{$\boldsymbol\kappa$} \\ \hline
15 & 0.893 & 0.893 & 0.893 & 0.893 & 0.786 \\ \hline
27 & 0.921 & 0.921 & 0.921 & 0.921 & 0.843 \\ \hline
32 & 0.954 & 0.954 & 0.954 & 0.954 & 0.907 \\ \hline
45 & 0.914 & 0.914 & 0.914 & 0.914 & 0.829 \\ \hline
51 & 0.936 & 0.936 & 0.936 & 0.936 & 0.871 \\ \hline
\textbf{AVERAGE} & 0.924 & 0.924 & 0.924 & 0.924 & 0.847 \\ \hline
\end{tabular}
\caption{Performance metrics of LightGBM on clade-assisted sentiment analysis}
\label{lightgbm}
\end{table}

LightGBM performed the best among all the models in the experiments. Its performance is quantitatively illustrated by Table~\ref{lightgbm}. The accuracy, precision, recall, and F1 scores are all equal for each seed respectively. The different metrics of the model ranges from 0.893 to 0.954. The model's mean accuracy, mean precision, mean recall, and mean F1 score are all equal to 0.924. The lowest value of $\kappa$ is 0.786, which is approaching a strong level of agreement. The highest value of $\kappa$ is 0.907, with an almost perfect level of agreement. The mean value of $\kappa$ is 0.847.\\

\begin{table}[H]
\begin{tabular}{|l|l|l|l|l|l|l|}
\hline
\textbf{neighbors} & \textbf{seed}    & \textbf{accuracy} & \textbf{precision} & \textbf{recall} & \textbf{f1 score} & \textbf{$\boldsymbol\kappa$} \\ \hline
10 & 15 & 0.857 & 0.857 & 0.857 & 0.857 & 0.714 \\ \hline
10 & 27 & 0.896 & 0.896 & 0.896 & 0.896 & 0.793 \\ \hline
10 & 32 & 0.929 & 0.929 & 0.929 & 0.929 & 0.857 \\ \hline
10 & 45 & 0.914 & 0.914 & 0.914 & 0.914 & 0.829 \\ \hline
10 & 51 & 0.889 & 0.889 & 0.889 & 0.889 & 0.779 \\ \hline
\textbf{AVERAGE} & - & 0.897 & 0.897 & 0.897 & 0.897 & 0.794 \\ \hline
\end{tabular}

\begin{tabular}{|l|l|l|l|l|l|l|}
\hline
\textbf{neighbors} & \textbf{seed}    & \textbf{accuracy} & \textbf{precision} & \textbf{recall} & \textbf{f1 score} & \textbf{$\boldsymbol\kappa$} \\ \hline
20 & 15 & 0.893 & 0.893 & 0.893 & 0.893 & 0.786 \\ \hline
20 & 27 & 0.861 & 0.861 & 0.861 & 0.861 & 0.721 \\ \hline
20 & 32 & 0.882 & 0.882 & 0.882 & 0.882 & 0.764 \\ \hline
20 & 45 & 0.921 & 0.921 & 0.921 & 0.921 & 0.843 \\ \hline
20 & 51 & 0.882 & 0.882 & 0.882 & 0.882 & 0.764 \\ \hline
\textbf{AVERAGE} & - & 0.888 & 0.888 & 0.888 & 0.888 & 0.776 \\ \hline
\end{tabular}

\caption{Performance metrics of K-nearest neighbors on clade-assisted sentiment analysis}
\label{knn}
\end{table}

The performance of the K-nearest neighbors algorithm on the test data is shown in Table~\ref{knn}. KNN with 10 neighbors performed better than 20 neighbors, in general. KNN with 10 neighbors has an average accuracy, average precision, average recall, and average F1 score of 0.897. Its average Cohen's kappa value is 0.794, which indicates a nearly strong level of agreement. As for KNN with 20 neighbors, average accuracy, average precision, average recall, and average F1 score are all equal---with a value of 0.888. The model's average $\kappa$ is 0.776, which means a moderate level of agreement.\\

\begin{table}[H]
\centering
\begin{tabular}{|l|l|l|l|l|l|}
\hline
\textbf{seed}    & \textbf{accuracy} & \textbf{precision} & \textbf{recall} & \textbf{f1 score} & \textbf{$\boldsymbol\kappa$} \\ \hline
15 & 0.686 & 0.686 & 0.686 & 0.686 & 0.371 \\ \hline
27 & 0.721 & 0.721 & 0.721 & 0.721 & 0.443 \\ \hline
32 & 0.721 & 0.721 & 0.721 & 0.721 & 0.443 \\ \hline
45 & 0.779 & 0.779 & 0.779 & 0.779 & 0.557 \\ \hline
51 & 0.786 & 0.786 & 0.786 & 0.786 & 0.571 \\ \hline
\textbf{AVERAGE} & 0.739 & 0.739 & 0.739 & 0.739 & 0.477 \\ \hline
\end{tabular}
\caption{Performance metrics of the Naive Bayes algorithm on clade-assisted sentiment analysis}
\label{naive-bayes}
\end{table}

The Naive Bayes' algorithm has the weakest performance metrics, especially in showing reliability. See Table~\ref{naive-bayes} to view the entire results. The accuracy, precision, recall, and F1 scores of the model have equal values across each of the seeds in the experiment---ranging from 68.6\% to 78.6\%, with an average value of 73.9\%. The model has displayed $\kappa$ values from 0.371 to 0.571. These values can be interpreted as having a minimal level of agreement to having a weak level of agreement. The model's mean $\kappa$ value is 0.477.\\

\begin{table}[H]
\centering
\begin{tabular}{|l|l|l|l|l|l|}
\hline
\textbf{seed}    & \textbf{accuracy} & \textbf{precision} & \textbf{recall} & \textbf{f1 score} & \textbf{$\boldsymbol\kappa$} \\ \hline
15 & 0.782 & 0.782 & 0.782 & 0.782 & 0.564 \\ \hline
27 & 0.782 & 0.782 & 0.782 & 0.782 & 0.564 \\ \hline
32 & 0.839 & 0.839 & 0.839 & 0.839 & 0.679 \\ \hline
45 & 0.796 & 0.796 & 0.796 & 0.796 & 0.593 \\ \hline
51 & 0.786 & 0.786 & 0.786 & 0.786 & 0.571 \\ \hline
\textbf{AVERAGE} & 0.797 & 0.797 & 0.797 & 0.797 & 0.594 \\ \hline
\end{tabular}
\caption{Performance metrics of the SVM model on clade-assisted sentiment analysis}
\label{svm}
\end{table}

The performance metrics of the support vector machine is shown in Table~\ref{svm}. The accuracy, precision, recall, and F1 scores of the SVM model ranges from 78.2\% to 83.9\% across all seeds. Its mean accuracy, mean precision, mean recall, and mean F1 score are all equal to 79.7\%. The minimum value of $\kappa$ is 0.564 and the model's maximum value of $\kappa$ is 0.679. The mean value of $\kappa$ is 0.594, which indicates an averagely weak level of agreement.

\section{Discussion}
The BERT topic model is able to extrapolate 25 distinct topics found in Tables~\ref{topics_1-6},~\ref{topics_7-12},~\ref{topics_13-18}, and~\ref{topics_19-25}. Through hierarchical clustering of the BERT topic embeddings, we are able to build a hierarchy of the 25 topics---found in Figure~\ref{hierarchy}. The green superclade clustered all topics about COVID-19 vaccine misinformation, while the blue superclade managed to group together COVID-19 misinformation, COVID-19 FAQs, and comments indicating desperation. All of the clades produced in Figure~\ref{hierarchy}, excluding the ``vaccination" topic in the yellow clade, are positive and negative topics. Particularly, the purple clade, containing religious comments, are the only positive topics. This indicates that aside from the purple clade, the majority of the hierarchy's topics are negative. It is also important to note that the hierarchical clustering model has precisely identified COVID-19 misinformation and COVID-19 vaccine misinformation, even if the detection of misinformation is not the objective of the study. This suggests COVID-19 misinformation is an emergent property of COVID-19 public opinion that does not require any analytical intention of discovery. The detection of COVID-19 misinformation can be an unsupervised task.\\

A temporal distribution analysis of the BERT topics produced reveals a fluctuating but upward trend for topics of vaccine administration (Figure~\ref{vaccine-administration}), vaccine concerns and complaints (Figure~\ref{vaccine-concerns}), and vaccine deaths (Figure~\ref{vaccine-deaths}). Based on the trigger events for these comments, Facebook posts containing comparisons to other countries increase the number of negative comments abruptly---suggesting xenocentrism. Furthermore, it is suggested that the high numbers of laugh reactions for these trigger events are not directed at the posts' contents themselves but at the Department of Health itself as an entity. This entails a public image issue.\\

The results of the vaccine brand reputation analysis via topic similarity (Figure~\ref{topic_similarity}) shows the public's association with the different brands of vaccines in the Philippines. Pfizer has low cosine similarities across its different similar topics. This indicates a good reputation since the topics belong to negative clades. The lower the cosine distances are, the better the public image is. It is notable that the negative comments are highly similar in terms of cosine distance, when considering the vaccine brands: Sputnik V, AstraZeneca, and Sinovac. Sputnik V's scores are heavily anchored on the supply shortage of the brand occurring in the country. AstraZeneca, on the other hand, is known to present the most adverse side effects, according to the data. Lastly, based on the comments, Sinovac is mostly associated by the public to the alleged vaccine-related deaths in the country and Sinovac's lesser efficacy in protecting an individual from COVID-19.\\

Lastly, the sentiment analysis reveals that the clade-assisted approach of sentiment classification offers generally favorable results. The worst performing model is the Naive Bayes' algorithm. It has an average accuracy, average precision, average recall, and average F1 value, all equal to 0.739. Its $kappa$ value is 0.477, indicating a weak level of agreement. It is probable that the kappa statistic is weak for this particular model of the data due to Naive Bayes' algorithm's assumption of independence among predictors. In reality, features of the Facebook comments are indeed related to one another. Meanwhile, LightGBM has performed the best among the five classifiers. The average accuracy, average precision, average recall, and average F1 value of the LightGBM model are all equal to 0.924. It also possesses a strong level of agreement when comparing the ground truth with the predictions ($\kappa$ = 0.847). This might be attributed to the gradient boosting framework used by LightGBM---which apparently makes XGBoost the second-best model, which is also based on the same framework. The performance of LightGBM suggests that the linguistic features present in the Facebook comments are sufficient to model public sentiment on Facebook posts made by the Department of Health Philippines, although different preprocessing techniques are applied beforehand.

\section{Conclusion}
The COVID-19 pandemic has radically changed the world. Over a year after the start of the outbreak, worldwide vaccine efforts against COVID-19 have begun. On the first day of March 2021, the Philippines started its vaccination program under the RESBAKUNA campaign. A recent study shows that vaccine hesitancy is a prevalent issue in the Philippines. The same concerns and complaints over the COVID-19 vaccines can also be found on social media, like Facebook. To understand and gauge national public opinion via the Facebook platform, the top comments from 50 Facebook posts of the official Facebook page of the Department of Health Philippines were scraped via Selenium and Python. The 50 Facebook posts are part of the ``\#RESBAKUNA" Facebook campaign by the agency. Next, a semi-supervised pipeline named Vaccine-related Exploratory Research via Topic Extraction, Brand Reputation Analysis, and Topic Emotions or VERTEBRATE is proposed. It is a pipeline that leverages topic modeling via multilingual BERT, analysis of vaccine brand reputation through cosine similarity of the BERT topic embeddings, hierarchical clustering of the BERT topic embeddings, and sentiment classification of the Facebook comments that are labeled automatically by the clades produced by the hierarchical clustering model.\\

The BERT topic model extracted 25 topics---2 of which are indeterminable, 2 of which are positive, and 21 topics are negative comments. These topics are precisely clustered through Ward's linkage hierarchical clustering. The model suggests that COVID-19 misinformation and COVID-19 vaccine misinformation are emergent properties of COVID-19 public opinion. Using supervised techniques to model such misinformation is unnecessary. In terms of the temporal distribution of these topics, the analysis suggests that negative comments drastically increase when the Department of Health Philippines posts about other countries. It can also be noted that the huge numbers of laugh reactions throughout the Facebook posts---without the presence of any humorous content---entail that the Department of Health Philippines has a public image issue. In this instance, people tend to react the way they do, not because of the posts but because of who posted them.\\

As for the brand reputation of the different COVID-19 vaccines, Pfizer and Moderna are doing relatively well---especially Pfizer's vaccine. Meanwhile, Sputnik V, AstraZeneca, and Sinovac suffer from negative associations. Additionally, clade-assisted sentiment classification effectively models public sentiment. The best-performing classifier, LightGBM, that is proposed in the study has managed to perform with 92.4\% accuracy. It also has a strong level of agreement in terms of Cohen's kappa statistic, with a value of 0.847.\\

Our present study highlights the persisting prevalence of COVID-19 vaccine misinformation in social media. Conspiracy beliefs and other forms of misinformation had been noted as a significant predictor of complete vaccine hesitancy \citep{al-sanafi_2021}. We noticed that DOH and its entities offered no responses to the comments posted by netizens under these infographics. Meta-analytic evidence suggests the importance of identifying misinformation most susceptible to correction, and engage experts in responding to misinformation \citep{walter_2020}. We recommend for DOH to form a social media team composed of health care professionals and interdisciplinary communication practitioners whose mandate is to respond to misinformation found in the comments of their posts. This engaging way of correcting false vaccine information will not only help quell vaccine doubts, but also hopefully improve the image of DOH among citizens.\\

Also, our present study suggests that instead of drawing aspirational sentiments from people, posting about the success of wealthier countries in terms of vaccination only intensifies Filipinos’ xenocentric tendencies to rationalize the poor COVID outcomes of the country. We recommend for DOH and other stakeholders involved in vaccine promotion to use narratives that are closer to home, like the diseases curbed by the expanded program on immunization in the Philippines. Using relatable narratives and storytelling had been indicated as effective means to combat anti-vaccine conspiracies \citep{lazic_2021}.\\

The VERTEBRATE pipeline effectively highlights the contents to avoid when posting on social media about COVID-19. Future work could include the implementation of an automatic labeling procedure for the topic model to further reduce manual effort.

\section*{Acknowledgements}
The authors would like to thank Nathaniel Oco for his invaluable insights, especially for suggesting the inclusion of Cohen's kappa statistic for evaluating the different classifiers and for proofreading the manuscript.

\newpage
 \bibliographystyle{elsarticle-num-names} 
 \bibliography{cas-refs}





\end{document}